\documentclass[final,5p,twocolumn]{elsarticle}

\usepackage[T1]{fontenc}
\usepackage{lmodern,microtype,amsmath,amssymb,amsthm}
\usepackage{etoolbox}
\usepackage{tikz}
\biboptions{numbers,sort&compress}
\usepackage[hidelinks]{hyperref}
\newtheorem{theorem}{Theorem}[section]
\newtheorem{lemma}[theorem]{Lemma}
\newtheorem{proposition}[theorem]{Proposition}

\newcommand{\ind}[1]{\mathbf{1}[#1]}
\newcommand{\botn}{\perp}
\journal{Information Processing Letters}

\makeatletter
\apptocmd{\ps@pprintTitle}{%
  \apptocmd{\@oddfoot}{\unskip}{}{}%
  \let\@evenfoot\@oddfoot
}{}{}
\makeatother

\hypersetup{
  pdftitle={A logarithm-free subexponential algorithm for minimum quartet inconsistency},
  pdfauthor={Chuang-Chieh Lin}
}

\begin{document}
\begin{frontmatter}
\title{A logarithm-free subexponential algorithm for minimum quartet inconsistency\tnoteref{label_title}}
\tnotetext[label_title]{This research was supported by National Science and Technology Council of Taiwan under grant no. NSTC 115-2221-E-019-045-}
\author[ntou]{Chuang-Chieh Lin\corref{cor1}}
\ead{josephcclin@mail.ntou.edu.tw}
\cortext[cor1]{Corresponding author.}
\address[ntou]{Department of Computer Science and Engineering,
National Taiwan Ocean University, Keelung City, Taiwan}

\begin{abstract}
We give a deterministic algorithm for complete, unweighted minimum quartet
inconsistency (MQI) with time and space $2^{O(k^{1/4})}+n^{O(1)}$, where~$n$ is the
number of taxa and~$k$ is the allowed number of quartet disagreements.
This removes the logarithmic factor from the exponent of the MQI bound
announced in 2009. The algorithm combines low-error anchoring, chromatic coding, and 
dynamic programming that accounts for disagreements with all input quartets, 
with a refined count of colorings and merging states. 
We also record a generic ETH-based lower bound. 
The optimality of the fourth-root exponent remains open. 
\end{abstract}
\begin{keyword}
Minimum quartet inconsistency \sep Parameterized algorithms \sep
Phylogenetic reconstruction \sep Chromatic coding \sep
Subexponential algorithms
\end{keyword}
\end{frontmatter}

\section{Introduction}
\label{sec:intro}

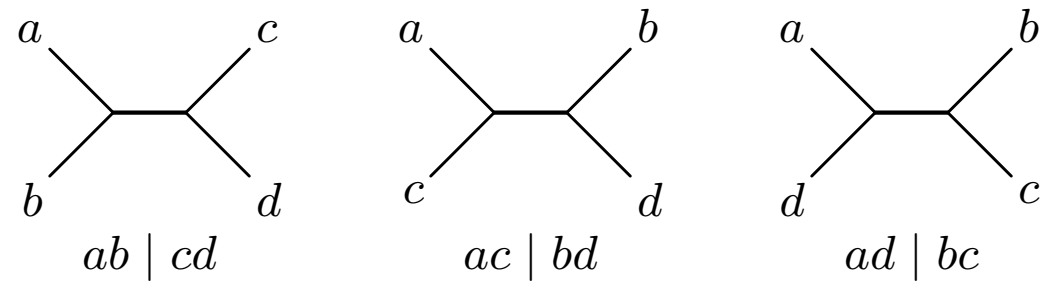
\begin{figure}[htbp]
  \centering
  \begin{tikzpicture}[
    x=0.60cm, y=0.60cm,
    line width=0.55pt,
    line cap=round, line join=round,
    every node/.style={font=\small, inner sep=1pt}
  ]
    \foreach \qshift/\qUL/\qLL/\qUR/\qLR in {
      0/a/b/c/d,
      4.2/a/c/b/d,
      8.4/a/d/b/c
    } {
      \begin{scope}[shift={(\qshift,0)}]
        \draw (-1.1,0.7) -- (-0.4,0) -- (-1.1,-0.7);
        \draw ( 1.1,0.7) -- ( 0.4,0) -- ( 1.1,-0.7);
        \draw[line width=0.8pt] (-0.4,0) -- (0.4,0);

        \node[above left]  at (-1.1, 0.7) {$\qUL$};
        \node[below left]  at (-1.1,-0.7) {$\qLL$};
        \node[above right] at ( 1.1, 0.7) {$\qUR$};
        \node[below right] at ( 1.1,-0.7) {$\qLR$};
        \node[below] at (0,-1.25)
          {$\qUL\qLL\mid\qUR\qLR$};
      \end{scope}
    }
  \end{tikzpicture}
  \caption{The three resolved quartet topologies on $\{a,b,c,d\}$.
    Each internal edge induces the indicated $2\mid2$ split.}
  \label{fig:quartet-topologies}
\end{figure}

Let $S$ be a set of $n\geq4$ taxa. An unrooted binary phylogenetic tree
on $S$ has its leaves labeled bijectively by $S$, and every internal vertex
has degree three. Such a tree on four taxa is called a \emph{resolved
quartet}. For $X=\{a,b,c,d\}$, its unique internal edge separates the leaves
into two pairs. We denote the topology by $ab\mid cd$ when these pairs are
$\{a,b\}$ and $\{c,d\}$. Figure~\ref{fig:quartet-topologies} shows the three
possible resolved topologies on $X$.

A \emph{quartet system} on $S$ is a collection of resolved quartets whose
leaf sets belong to $\binom S4$. We call the system \emph{full} if each
four-element subset of $S$ is represented by exactly one quartet.
We identify a full system with a function $Q$, writing $Q(X)$ for the
selected topology on $X\in\binom S4$. Throughout this note, a
\emph{complete} quartet input means a full quartet system in this sense. 
Note that the chosen topologies need not all be displayed by a single tree.

For an unrooted binary phylogenetic tree $T$ on $S$ and $X\in\binom S4$,
let $T|_X$ be the tree obtained from the minimal connected subtree of $T$
spanning $X$ by suppressing all degree-two vertices. We say that $T$
\emph{displays} a quartet $q$ on $X$ if $T|_X=q$. The induced full quartet
system $Q_T$ is defined by $Q_T(X)=T|_X$ for every $X\in\binom S4$.
A full system $Q$ is \emph{compatible} if some tree $T$ displays every
quartet selected by $Q$, equivalently, if $Q=Q_T$.

For a complete input $Q$ and a candidate tree $T$, define
\[
 d(Q,T)=\sum_{X\in\binom S4}
 \mathbf{1}[Q(X)\ne Q_T(X)],
\]
where $\mathbf{1}[P]$ equals one when the statement $P$ is true and zero
otherwise. Minimum quartet inconsistency (MQI) asks whether
$d(Q,T)\leq k$ for some unrooted binary phylogenetic tree $T$ on $S$.
Thus $k$ counts quartet entries that must be \emph{changed}, rather than taxa,
tree rearrangements, or missing entries. Inputs with at most three taxa
have no quartet entries and can be handled separately.

\subsection{Related work}

Gramm and Niedermeier~\cite{GN03} established fixed-parameter tractability
with an $O(4^k n+n^4)$ algorithm. Chang, Lin, and Rossmanith~\cite{CLR10}
improved the branching guarantees and gave $O^*((1+\varepsilon)^k)$
algorithms\footnote{The notation $O^*$ suppresses factors polynomial in the input size.} 
for each fixed $\varepsilon>0$. Wu, You, and Lin~\cite{WYL06}
obtained polynomial-time algorithms when $k=O(n)$ with a fixed implied
constant. Their low-error-anchor observation~\cite[Lemma~2.2]{WYL06}
is a starting point of our construction.

Alon, Lokshtanov, and Saurabh's chromatic-coding
framework~\cite{ALS09} led to the announcement of an MQI algorithm with
running time $2^{O(k^{1/4}\log k)}n^{O(1)}$~\cite[p.~4]{S09};
the joint attribution is also stated in~\cite{S11}.
Thus subexponential fixed-parameter tractability and the fourth-root
power of~$k$ were already known.

Guillemot and Mnich~\cite{GM13} give a
$2^{O(p^{1/3}\log p)}+O(n^4)$ algorithm for dense rooted triplet
inconsistency, where $p$ is the triplet-error budget.
Their weak-coloring construction and tree-position product states
\cite[Lemmas~8 and~9]{GM13} are direct methodological predecessors of our
algorithm. 
Bryant and Steel~\cite{BS01} developed dynamic programming for quartet 
optimization over trees whose splits belong to a prescribed collection. 
For each cluster, their partial score includes quartets lying entirely 
within the cluster, together with quartets having three leaves inside and 
one outside. We use these established
coloring, merging, and quartet-scoring ingredients.

Other quartet results address different models. 
Lafond and Scornavacca~\cite{LS19} study weighted quartet consensus, 
where quartet weights are their frequencies among input trees on the same taxon set. 
Their fixed-parameter-tractable algorithms use parameters counting four-taxon sets with tied most-frequent 
topologies and uniquely most-frequent quartets that the solution is allowed to reject. 
Hirai and Iwamasa~\cite{HI22} give polynomial-time compatibility algorithms for multi-partite quartet
systems. Recent work on reconstruction from random noisy
samples~\cite{ACLM26} and sublinear-time consistency-versus-farness-from-consistency
testing~\cite{Lin26} concerns different objectives or access models from
the exact complete-input repair problem considered here.

\subsection{Our result and approach}

\begin{theorem}\label{thm:main}
Complete unweighted MQI admits a deterministic algorithm returning a witnessing
tree on a yes-instance with time and space at most
\[
 2^{O(k^{1/4})}+n^{O(1)}.
\]
The hidden constants are independent of~$n$ and~$k$.
\end{theorem}

This removes the logarithmic factor from the previously announced
$2^{O(k^{1/4}\log k)}n^{O(1)}$ guarantee while retaining its fourth-root
power. To the best of our knowledge, this combined logarithm-free MQI
guarantee has not appeared in the publicly available literature.

Our contribution combines low-error anchoring with dynamic programming that 
accounts for disagreements with all input quartets. 
The improvement comes from counting repair sets using binomial sums and 
bounding the number of merging states using the product of the color-class sizes.
We call a chosen reference taxon~$a\in S$ an \emph{anchor}. 
Quartets containing $a$ are converted into rooted triplets on the remaining taxa. 
For any target tree with at most~$k$ quartet disagreements, 
an averaging argument guarantees an anchor contained in
at most~$p=\lfloor4k/n\rfloor$ disagreeing quartets.
Removing the anchor from their leaf sets gives a
3-uniform error hypergraph on $S\setminus\{a\}$. Let $h$ be the smallest power of two satisfying
$h^3\geq\max\{1,p\}$. A universal weak-coloring family contains a coloring under which
no error triple is monochromatic. Each color class then yields
a rooted restriction of the target tree, and these restrictions
can be merged while scoring all input quartets. 
Instead of replacing the repair-set count and the class-size product by
$n^{O(h)}$, we retain their actual sizes and apply elementary binomial and
AM--GM estimates. The resulting bound is
$2^{O(h\log(1+n/h))}n^{O(1)}$ (see Proposition~\ref{prop:anchored}, which is 
established by combining the coloring-family bound in Lemma~\ref{lem:color} with 
the merging time analysis in Lemma~\ref{lem:merge}). 
For positive~$p$, we have $h=O(p^{1/3})$ and $np\leq4k$,
so $nh^3=O(np)=O(k)$. Lemma~\ref{lem:entropy} therefore gives
$h\log(1+n/h)=O(k^{1/4})$, removing the logarithmic factor
from the exponent. The case $p=0$ is handled separately
by reconstruction and verification.

The remainder of this paper is organized as follows.
Section~\ref{sec:anchor_coloring} develops the anchor argument and the
universal weak-coloring family. Section~\ref{sec:quartet_optimization}
presents the quartet-scoring recurrence and the anchored algorithm, and
Section~\ref{sec:entropy} proves the logarithm-free running-time bound.
Section~\ref{sec:lower} gives the generic ETH consequences and a criterion
for stronger lower bounds. Section~\ref{sec:concluding} concludes with
directions for future work.

\section{Anchors and universal colorings}
\label{sec:anchor_coloring}

\subsection{Low-error anchors}

For a fixed target tree $T^\star$ on $S$, define its set of
erroneous four-taxon sets by
\[
 E=
 \left\{
 X\in\binom{S}{4}: Q(X)\ne T^\star|_X
 \right\}.
\]
Thus $|E|=d(Q,T^\star)$, and $T^\star$ is feasible if
$|E|\leq k$. For each $a\in S$, let $e_a=|\{X\in E:a\in X\}|$ 
be the number of erroneous quartets containing $a$.
The following averaging observation is due to Wu, You,
and Lin~\cite[Lemma~2.2]{WYL06}. We retain its dependence
on~$k$ and~$n$ explicitly.

\begin{lemma}\label{lem:anchor}
If $|E|\leq k$, there exists an anchor $a\in S$ such that
\[
 e_a\leq p:=\left\lfloor\frac{4k}{n}\right\rfloor.
\]
\end{lemma}

\begin{proof}
Each erroneous quartet contains four taxa, so
\[
 \sum_{a\in S}e_a=4|E|\leq4k.
\]
By averaging, there exists $a\in S$ such that
\[
 e_a
 \leq\frac{1}{n}\sum_{b\in S}e_b
 =\frac{4|E|}{n}
 \leq\frac{4k}{n}.
\]
Since $e_a$ is an integer, this implies
$e_a\leq\lfloor4k/n\rfloor=p$.
\end{proof}

For a fixed anchor $a$, let $L=S\setminus\{a\}$.
Transform every input quartet $xy\mid za$ containing $a$
into the rooted triplet $xy\mid z$ on $L$, and denote the
resulting complete rooted triplet system by $R_a$.

Deleting $a$ from an unrooted binary tree on $S$ and
rooting at its former neighbor gives a rooted binary tree
on $L$. Under this correspondence, the unrooted tree
displays $xy\mid za$ if and only if the rooted tree
displays $xy\mid z$. Conversely, attaching $a$ as a new
leaf adjacent to the root of a rooted binary tree on $L$
recovers an unrooted binary tree on $S$.

Let $V_a^\star$ denote the rooted tree obtained from
$T^\star$ by this operation. Exactly $e_a$ triplets of
$R_a$ are not displayed by $V_a^\star$. Consequently,
for the anchor guaranteed by Lemma~\ref{lem:anchor},
at most $p$ input rooted triplets disagree with the
target rooted tree.

Instances with $n\leq3$ have no quartet entries.
If $k\geq\binom{n}{4}$, any binary tree on $S$ suffices.
These cases can therefore be handled directly.

If $n>4k$, then $p=0$. For each anchor $a$, test whether
$R_a$ is consistent. If it is, reconstruct its rooted
tree, attach $a$ to the root, and check whether the
resulting unrooted tree $T$ satisfies $d(Q,T)\leq k$.
Accept if any candidate passes this check, and otherwise
reject. On every yes-instance, Lemma~\ref{lem:anchor}
provides an anchor with $e_a=0$, for which reconstruction
recovers the target tree. Thus this procedure is exact
and runs in polynomial time.

Rooted reconstruction can use Aho et al.'s
BUILD algorithm~\cite{ASSU81}, followed by direct
verification. A consistent complete rooted triplet system determines
a unique rooted binary tree. For a one-taxon set, this
tree consists of a single vertex serving as both root
and leaf. For a two-taxon set, it consists of a root
whose two children are the labeled leaves.
These cases require no triplet constraints or
reconstruction procedure.

For the remaining analysis, we may therefore assume
$n\geq4$, $1\leq k<\binom{n}{4}$ and $n\leq4k$.

\subsection{A universal family for anchored errors}

For the fixed target tree $T^\star$ and an anchor $a$,
define the associated \emph{error hypergraph}
$H_a=(L,E_a)$, where $L=S\setminus\{a\}$ and
\[
 E_a=
 \left\{
 X\in\binom{L}{3}:
 Q(X\cup\{a\})\ne T^\star|_{X\cup\{a\}}
 \right\}.
\]
Each hyperedge is the three-element leaf set of an input
rooted triplet in $R_a$ that is not displayed by
$V_a^\star$. The hyperedge itself carries no tree
topology. Removing $a$ gives a bijection between the
erroneous four-taxon sets containing $a$ and the
hyperedges of $H_a$, so $|E_a|=e_a$.
If $T^\star$ is feasible, Lemma~\ref{lem:anchor}
therefore guarantees at least one anchor with
$|E_a|\leq p$. This bound need not hold for every anchor.

\paragraph{Remark} The error hypergraph depends on the unknown target tree
and is used only in the analysis. We construct a family
of colorings that works for every hypergraph with the
specified number of vertices and at most $p$ hyperedges,
without knowing which hyperedges occur.

A \emph{proper weak vertex coloring} of a hypergraph
assigns colors to its vertices so that every hyperedge
contains at least two distinct colors. Thus, for a
3-uniform hypergraph, a coloring is proper weak if no
hyperedge has all three vertices of the same color.

The purpose of coloring is to partition the taxa into
classes whose anchored triplets agree with the target.
A proper weak coloring of $H_a$ ensures that every
error hyperedge meets at least two color classes.
Thus no erroneous rooted triplet is contained entirely
within a single class, allowing each class's rooted
restriction to be reconstructed correctly.
This also explains why weak coloring suffices:
the three vertices of an error hyperedge need not
receive three distinct colors.

The following construction adapts the proof
of~\cite[Lemma~9]{GM13} to 3-uniform hypergraphs.
We bound its construction time in terms of both the
number $m$ of vertices and the upper bound $p$ on the
number of hyperedges.

\begin{lemma}\label{lem:color}
Let $V$ be an $m$-element set with $m\geq1$, let $p$
be a nonnegative integer, and let $h$ be the smallest
power of two satisfying $h^3\geq\max\{1,p\}$.
A family of vertex colorings of $V$, each using at most
$2h$ colors, can be constructed deterministically in time
\[
 2^{O(h\log(1+m/h))}m^{O(1)}.
\]
Every 3-uniform hypergraph on $V$ with at most $p$
hyperedges is properly weakly colored by at least one
coloring in the family.
\end{lemma}

\begin{proof}
If $h\geq m$, use the single injective coloring, which
assigns a distinct color to every vertex. This coloring
uses at most $h$ colors and properly weakly colors every
3-uniform hypergraph on $V$.

Now suppose that $h<m$. Let $q$ be the smallest power
of two with $q\geq m$, and assign distinct field elements
$x_v\in\mathbb F_q$ to the vertices $v\in V$.
Then $m\leq q<2m$. Since $h$ and $q$ are powers of two
and $h<q$, we may view $\mathbb F_q$ as a vector space
over $\mathbb F_2$ and fix a surjective
$\mathbb F_2$-linear projection
\[
 \Pi:\mathbb F_q\to
 \mathbb F_2^{\log_2 h}.
\]
Identify the $h$ possible outputs of $\Pi$ with the
initial colors $1,2,\ldots,h$.

For each coefficient vector
$(b_0,b_1,b_2)\in\mathbb F_q^3$, define an initial
coloring by
\[
 g(v)=\Pi(b_0+b_1x_v+b_2x_v^2).
\]
Thus one coefficient vector determines one coloring of
all vertices. When $h=1$, the projection has only one
possible output, so every coefficient vector gives the
same constant coloring.

A uniformly chosen coefficient vector gives three-wise
independent uniform colors. To see this, fix any three
distinct vertices $v_1,v_2,v_3$, and write
$x_i=x_{v_i}$ and $y_i=b_0+b_1x_i+b_2x_i^2$ for
$i=1,2,3$. Then
\[
 \begin{pmatrix}y_1\\y_2\\y_3\end{pmatrix}
 =
 \begin{pmatrix}
  1&x_1&x_1^2\\
  1&x_2&x_2^2\\
  1&x_3&x_3^2
 \end{pmatrix}
 \begin{pmatrix}b_0\\b_1\\b_2\end{pmatrix}.
\]
The determinant of this Vandermonde matrix is
\[
 (x_2-x_1)(x_3-x_1)(x_3-x_2)\ne0,
\]
since the $x_i$ are distinct. Hence every prescribed
value vector $(y_1,y_2,y_3)\in\mathbb F_q^3$
corresponds to exactly one coefficient vector.
Uniformly choosing the coefficient vector therefore
makes all $q^3$ value vectors equally likely.

Each color has exactly $q/h$ preimages under $\Pi$.
Consequently, for any prescribed colors $c_1,c_2,c_3$,
\[
 \begin{aligned}
 &\Pr[g(v_1)=c_1,\ g(v_2)=c_2,\ g(v_3)=c_3] = \frac{(q/h)^3}{q^3} = \frac{1}{h^3}.
 \end{aligned}
\]
This establishes three-wise independence and uniformity.
In particular, any fixed hyperedge is monochromatic
with probability $h\cdot h^{-3}=h^{-2}$.

Fix an arbitrary 3-uniform hypergraph
$H=(V,\mathcal E)$ with $|\mathcal E|\leq p$.
By linearity of expectation, the expected number of
monochromatic hyperedges is
\[
 \frac{|\mathcal E|}{h^2}
 \leq\frac{p}{h^2}
 \leq h.
\]
Thus at least one coefficient vector defines a coloring
$g$ with at most $h$ monochromatic hyperedges.
Choose one vertex from each such hyperedge, and let
$Z$ be the set of chosen vertices. Then $|Z|\leq h$.

Give each vertex of $Z$ its own fresh color, distinct
from all initial colors and from the fresh colors
assigned to the other vertices of $Z$.
Every previously monochromatic hyperedge meets $Z$.
Moreover, every hyperedge meeting $Z$ contains a vertex
whose fresh color appears nowhere else, so that
hyperedge is non-monochromatic after recoloring.
Hyperedges disjoint from $Z$ retain their original
colors and were already non-monochromatic.
The resulting coloring therefore properly weakly
colors $H$ and uses at most $h+|Z|\leq2h$ colors.

To construct the family without knowing $H$, enumerate
all $q^3$ coefficient vectors and construct their
corresponding initial colorings. For each initial
coloring, enumerate all subsets $Z\subseteq V$ with
$|Z|\leq h$ and perform the recoloring above.
Fix an ordering of $V$ and assign the fresh colors
$h+1,h+2,\ldots,h+|Z|$ to the vertices of $Z$ in that order.

The resulting family contains at most~$q^3\sum_{j=0}^{h}\binom{m}{j}$
colorings, counting repetitions. By the preceding
argument, it contains a proper weak coloring for every
3-uniform hypergraph on $V$ with at most $p$ hyperedges.
Though different coefficient vectors may produce the same
initial coloring, this does not affect the averaging
argument, which uses the uniform distribution over
coefficient vectors.

We next bound the number of possible repair sets $Z$.
First suppose that $h\leq m/2$, and set $x=h/m$.
Since $0<x\leq1$, we have $x^j\geq x^h$ for every
$0\leq j\leq h$. Therefore,
\[
 \begin{aligned}
 x^h\sum_{j=0}^{h}\binom{m}{j}
 &\leq\sum_{j=0}^{h}\binom{m}{j}x^j\\
 &\leq\sum_{j=0}^{m}\binom{m}{j}x^j\\
 &=(1+x)^m\\
 &\leq e^{mx}=e^h.
 \end{aligned}
\]
Here we used the binomial theorem, the inequality
$1+x\leq e^x$, and the identity $mx=h$.
Dividing by $x^h=(h/m)^h$ gives
\[
 \sum_{j=0}^{h}\binom{m}{j}
 \leq\frac{e^h}{(h/m)^h}
 =\left(\frac{em}{h}\right)^h.
\]
Since $m/h\geq2$, we have $e<3\leq1+m/h$ and
$m/h\leq1+m/h$. Hence
\[
 \begin{aligned}
 \left(\frac{em}{h}\right)^h
 &\leq\left(1+\frac{m}{h}\right)^{2h}\\
 &=2^{2h\log_2(1+m/h)}.
 \end{aligned}
\]

If $m/2<h<m$, then $m<2h$, and
\[
 \sum_{j=0}^{h}\binom{m}{j}
 \leq\sum_{j=0}^{m}\binom{m}{j}
 =2^m
 <2^{2h}.
\]
Since $\log_2(1+m/h)>1$, the same bound
$2^{O(h\log(1+m/h))}$ holds in this case as well.

Finally, $q<2m$, so enumerating the coefficient vectors
contributes only a polynomial factor in~$m$.
Constructing the finite field, evaluating each initial
coloring, and producing each recolored assignment also
require only polynomial overhead in~$m$.
The case $h\geq m$ requires only the single injective
coloring and is polynomial.
This proves the claimed construction-time bound.

\end{proof}

\paragraph{Remark} The dependence on the ratio $m/h$ rather 
than the weaker estimate 
$\sum_{j=0}^{h}\binom{m}{j}\leq \sum_{j=0}^h m^j\leq m^{O(h+1)}$ 
will be used in Section~\ref{sec:entropy} to obtain the logarithm-free
$O(k^{1/4})$ exponent.

For each anchor $a$, construct the family from
Lemma~\ref{lem:color} on $L=S\setminus\{a\}$, with
$m=n-1$ and $p=\lfloor4k/n\rfloor$.
For each coloring in the family, let
$C_1,C_2,\ldots,C_t$ be its nonempty color classes, where
$t\leq2h$. Test each restricted rooted triplet system
$R_a|_{C_i}$ for consistency. Reject the coloring if any
restriction is inconsistent. Otherwise, reconstruct its
rooted tree~$U_i$.

To see why this preserves a feasible target, suppose
that $d(Q,T^\star)\leq k$. Choose an anchor with
$|E_a|\leq p$, as guaranteed by Lemma~\ref{lem:anchor}.
Lemma~\ref{lem:color} then provides a coloring in the
family that properly weakly colors $H_a$.
No hyperedge of $H_a$ is contained in a single color
class. Thus every rooted triplet of $R_a|_{C_i}$ is
displayed by $V_a^\star|_{C_i}$.
Since these restricted triplet systems are complete,
reconstruction yields $U_i=V_a^\star|_{C_i}$ for every~$i$.
The same statement holds for classes of size one or two,
whose rooted trees are determined directly.
Consequently, at least one anchor and one coloring
produce exactly the rooted restrictions of the
feasible target.

\paragraph{Remark}
We do not require the quartet system~$Q|_{C_i}$ to be
consistent. Even when a coloring properly weakly
colors~$H_a$, input quartets entirely within~$C_i$
may disagree with~$T^\star$, because they do not
contain the anchor.
The class tree~$U_i$ is inferred exclusively from
the anchored rooted triplets.
Disagreements with all input quartets are evaluated
by the optimization step in
Section~\ref{sec:quartet_optimization}.

\section{Optimizing the full quartet objective}
\label{sec:quartet_optimization}

\subsection{Quartet charging with a permanent outgroup}

For distinct taxa $x,y,z,w$, write
\[
 w_Q(xy\mid zw)=\ind{Q(\{x,y,z,w\})\ne xy\mid zw}.
\]
Let $A$ and $B$ be the leaf sets of the two child subtrees
of an internal node, and let $O=S\setminus(A\cup B)$
denote the set of outside taxa. Define
\begin{align}
 c_Q(A,B,O)={}&
 \sum_{\substack{\{x,x'\}\subseteq A\\\{y,y'\}\subseteq B}}
 w_Q(xx'\mid yy')\notag\\
 &+\sum_{\substack{\{x,x'\}\subseteq A\\y\in B,\ z\in O}}
 w_Q(xx'\mid yz)\notag\\
 &+\sum_{\substack{x\in A,\ \{y,y'\}\subseteq B\\z\in O}}
 w_Q(xz\mid yy').
 \label{eq:charge}
\end{align}
The three lines correspond to distributions $(2,2,0)$, $(2,1,1)$, and
$(1,2,1)$ among $A,B,O$. An explicit computation costs $O(n^4)$ time.

\begin{lemma}\label{lem:charging}
Let $V$ be a rooted binary tree on $L=S\setminus\{a\}$, and let $T$ result
from attaching $a$ to its root. For each internal node $v$ of $V$, let $A_v,B_v$
be its two child leaf sets, and put $O_v=S\setminus(A_v\cup B_v)$. Then
\[
 d(Q,T)=\sum_{v\text{ internal}}c_Q(A_v,B_v,O_v).
\]
Each four-taxon set $X\in\binom{S}{4}$ contributes exactly
one term to the sum, namely $\mathbf{1}[Q(X)\ne T|_X]$.
\end{lemma}
\begin{proof}
We show that each four-taxon set $X\in\binom{S}{4}$ contributes
exactly one term to the sum, and that this term equals
$\mathbf{1}[Q(X)\neq T|_X]$.

First, suppose that $X = \{a, x, y, z\}\subseteq S$. Let $u$ be 
the lowest common ancestor of~$x$, $y$, $z$ in~$V$. 
Because $V$ is binary, these three leaves split two-to-one 
between the children of~$u$. Suppose that $x,y\in A_u$ and $z\in B_u$. 
Since the anchor is outside every rooted subtree of~$V$, we have~$a\in O_u$. 
Consequently, the second sum in~\eqref{eq:charge} charges 
$w_Q(xy\mid za)$, which is precisely the disagreement with the quartet 
displayed by~$T$. Note that this charge is unique. Indeed, 
a proper descendant of~$u$ contains at most two of~$x$, $y$, and $z$, 
while a proper ancestor has all three in one child subtree and 
an incomparable node contains none of them. 
The case with the pair in $B_u$ similarly follows. 

Second, suppose that $X\subseteq L$, and let $v$ be the lowest
common ancestor of its four leaves in~$V$. The two child
subtrees of $v$ contain either two leaves each or one and
three leaves.

In the two-to-two case, $X$ contributes through the first
sum in~\eqref{eq:charge} at~$v$. The two pairs in the
child subtrees determine the split displayed by $T|_X$.
To show that this contribution is unique, consider any
other internal node~$u$.
If $u$ is a proper descendant of~$v$, its rooted subtree
lies within one child subtree of~$v$ and therefore
contains at most two leaves of~$X$.
If $u$ is a proper ancestor of~$v$, all four leaves of~$X$ 
lie in the child subtree of~$u$ that contains $v$.
Otherwise, $u$ is neither an ancestor nor a descendant
of~$v$, so their rooted subtrees are disjoint.
Since all leaves of~$X$ lie below~$v$, none lies below~$u$
in this last case.
Each sum in~\eqref{eq:charge} requires at least three
leaves of~$X$ below the node, with at least one in each
child subtree. None of the other nodes satisfies both
requirements. Hence $X$ contributes only at~$v$.

In the one-to-three case, let $Y$ be the three leaves
of $X$ in one child subtree of $v$, let $z$ be the
fourth leaf, and let $u$ be the lowest common ancestor
of $Y$. Since $Y$ lies in one child subtree of~$v$,
$u$ is a proper descendant of~$v$.
The leaves of $Y$ divide two-to-one between the child
subtrees of $u$, while $z$ lies in~$O_u$.
Thus $X$ contributes through exactly one of the last
two sums in~\eqref{eq:charge} at~$u$.
The two leaves in the same child subtree form one side
of the split displayed by $T|_X$.
To show that this contribution is unique, first observe
that the distribution at $v$ is $(3,1,0)$ or~$(1,3,0)$,
so no term for $X$ is included there.
The subtree rooted at any proper descendant of~$u$
contains at most two leaves of~$X$.
At any node strictly between $u$ and~$v$, all three
leaves of~$Y$ lie in a single child subtree, while~$z$ 
lies outside that node's subtree.
At any proper ancestor of~$v$, all four leaves of~$X$
lie in a single child subtree.
Finally, any node that is neither an ancestor nor a
descendant of~$u$ and is distinct from~$u$ has a rooted
subtree disjoint from the subtree rooted at~$u$.
Its subtree therefore contains no leaves of~$Y$ and
can contain at most the single leaf $z$ from~$X$.
Consequently, no node other than~$u$ has any of the
distributions $(2,2,0)$, $(2,1,1)$, or $(1,2,1)$
required by~\eqref{eq:charge}.
Hence $X$ contributes only at~$u$.

In every case, the unique term for~$X$ evaluates its
disagreement with the topology displayed by~$T$.
Summing over all $X\in\binom{S}{4}$ gives
\[
 \sum_{v\text{ internal}}c_Q(A_v,B_v,O_v)
 =
 \sum_{X\in\binom{S}{4}}
 \mathbf{1}[Q(X)\ne T|_X]
 =
 d(Q,T).
\]
\end{proof}

For the dynamic program below, consider a rooted subtree
on a cluster $C\subseteq L$, attached to the rest of the
tree only at its root. Its partial score includes the
disagreement costs of all quartets with at least three
leaves in $C$.
Quartets entirely within $C$ are determined by this
subtree. For distinct $x,y,z\in C$ and
$w\in S\setminus C$, if the subtree displays the rooted
triplet $xy\mid z$, then every completion preserving
$C$ as a cluster displays the quartet $xy\mid zw$.
Thus the costs of quartets with three leaves in $C$
and one outside depend on the rooted subtree and the
identities of the outside leaves through $Q$, but not
on how the rest of the tree is arranged.
A state on $C$ can therefore minimize this partial
score over its admissible rooted subtrees without
specifying an outside completion.
Quartets with fewer than three leaves in $C$ are
excluded from this partial score and counted elsewhere
in the full recurrence, as guaranteed by
Lemma~\ref{lem:charging}.

This partial-score formulation follows Bryant and
Steel~\cite[Section~2.3, Eqs.~(6)--(12)]{BS01}.
The binary recurrence in~\eqref{eq:product} in 
Section~\ref{subsec:product_states} specializes it to 
our setting.

\subsection{Product states for fixed rooted restrictions}
\label{subsec:product_states}

Suppose the disjoint nonempty classes $C_1,C_2,\ldots,C_t$ partition $L$, with
fixed rooted binary trees $U_1,U_2,\ldots,U_t$. A state $\pi=(u_1,u_2,\ldots,u_t)$ chooses, for each $i$, either
a node $u_i$ of $U_i$ or the special symbol $\botn$.
A chosen node represents its descendant leaf set, whereas
$u_i=\botn$ represents the empty set, indicating that the
state contains no leaves from~$C_i$. 
Let $C(\pi)$ be the union of the descendant leaf sets, interpreting $\botn$
as empty. This is the tree-position representation of~\cite[Lemma~8]{GM13}.

Next, we describe how the dynamic program splits its current
leaf set into two child subtrees while preserving the prescribed
class trees $U_1,U_2,\ldots,U_t$. For a state
$\pi=(u_1,u_2,\ldots,u_t)$, write its two child states as
$\pi_1=(v_1,v_2,\ldots,v_t)$ and $\pi_2=(w_1,w_2,\ldots,w_t)$,
representing the left and right subtrees, respectively.
For each coordinate $i$, we choose the ordered pair $(v_i,w_i)$.
If $u_i=\botn$, this class contributes no leaves to the current
state, so the only choice is $(v_i,w_i)=(\botn,\botn)$.
If $u_i\neq\botn$, the permitted choices are
\[
\begin{array}{c|cc}
 \text{choice} & v_i & w_i\\ \hline
 \text{entirely left}  & u_i & \botn\\
 \text{entirely right} & \botn & u_i\\
 \text{split}          & x_i & y_i\\
 \text{reversed split} & y_i & x_i
\end{array}
\]
Here the first two choices place all leaves descending from
$u_i$ into one child subtree. The last two choices are available
only when $u_i$ is an internal node with children $x_i,y_i$.
If the leaves represented by $u_i$ occur in both child subtrees,
their division must follow these two prescribed child
subtrees, in either order, to preserve the class tree.
Thus a null coordinate has one choice, a leaf coordinate
has two choices, and an internal-node coordinate has four
choices.

Choosing one permitted pair for each of the $t$ coordinates
therefore gives at most $4^t$ candidate transitions.
For each candidate $(\pi_1,\pi_2)$, put
$A=C(\pi_1)$ and $B=C(\pi_2)$. The coordinate rules ensure
that $A$ and $B$ are disjoint and that $A\cup B=C(\pi)$.
We retain only transitions for which both $A$ and $B$
are nonempty.

Define $D(\pi)=0$ if $|C(\pi)|\leq1$, and otherwise set
\begin{equation}
\begin{aligned}
 D(\pi)=\min_{(\pi_1,\pi_2)}\bigl\{&
 D(\pi_1)+D(\pi_2)\\
 &{}+c_Q(A,B,S\setminus C(\pi))\bigr\},
\end{aligned}
\label{eq:product}
\end{equation}
where the minimum is taken over the retained transitions.
The first two terms account for costs within the child
states, while the last term accounts for quartet costs
charged at the current split.

Since both child leaf sets are nonempty and partition~$C(\pi)$, 
each child state contains strictly fewer leaves
than its parent. Hence the recurrence is acyclic and can
be evaluated in increasing order of~$|C(\pi)|$.
The root state is $(r_1,r_2,\ldots,r_t)$, where $r_i$ is the
root of~$U_i$. Its leaf set is all of~$L$.

\begin{lemma}\label{lem:merge}
At its root state, \eqref{eq:product} computes the minimum full quartet
score among all trees whose rooted restriction to each $C_i$ is $U_i$.
Writing $m=n-1$ and $n_i=|C_i|$, its time is at most
\[
 n^{O(1)}4^t\prod_{i=1}^t2n_i,
\]
with space at most $n^{O(1)}\prod_{i=1}^t2n_i$. Moreover, it can return a tree.
\end{lemma}
\begin{proof}
For a nonempty state $\pi=(u_1,u_2,\ldots,u_t)$, call a rooted
binary tree $R$ on~$C(\pi)$ feasible if, for every non-null
coordinate~$u_i$, its restriction to~$C_i\cap C(\pi)$ is
the rooted subtree of $U_i$ at~$u_i$.

We first show that the permitted transitions describe exactly
the possible root splits of feasible trees. Consider a feasible
tree $R$ with at least two leaves. A null coordinate contributes
no leaves to either child, so its child coordinates must both
be $\botn$. For a non-null coordinate $u_i$, if all its
represented leaves lie in one child of $R$, the child coordinates
are $(u_i,\botn)$ or $(\botn,u_i)$. Otherwise, both children
contain represented leaves from this class. The root of $R$
then survives when restricting to these leaves and suppressing
unary nodes. Consequently, $u_i$ must be internal, and the two
child restrictions must be the subtrees rooted at its children
$x_i,y_i$. This gives $(x_i,y_i)$ or $(y_i,x_i)$.
Thus every feasible tree determines a permitted transition
whose child trees are feasible for their respective states.

Conversely, joining two feasible child trees under a new root
according to a permitted transition preserves every prescribed
class restriction. Indeed, each represented class either remains
entirely in one child or is joined along its prescribed split.
The requirement that both child leaf sets be nonempty ensures
that the resulting tree is binary. Hence the recurrence
represents all feasible trees and produces only feasible trees.

For a feasible tree at a state $\pi$, define its partial score
as the sum of the quartet charges at its internal nodes,
with each outside set taken relative to~$S$.
At a root split, this partial score is the sum of the two child
partial scores and the charge at the root. The base case has
score zero. Moreover, every permitted transition has two
nonempty child leaf sets strictly smaller than $C(\pi)$.
Induction on $|C(\pi)|$ therefore shows that the recurrence
computes the minimum partial score over all trees feasible
for $\pi$. At the root state, whose leaf set is
$L=S\setminus\{a\}$, Lemma~\ref{lem:charging} identifies this
sum with the full quartet-disagreement score of the
corresponding tree on $S$. This proves correctness.

For each non-base state, record one pair of child states
attaining the minimum. Recursively following these choices
from the root state reconstructs an optimal feasible rooted
tree on $L$. Reattaching the anchor $a$ yields the corresponding
unrooted tree on $S$.

It remains to bound the running time. Let $n_i=|C_i|$, so that
$\sum_{i=1}^{t}n_i=n-1$. The rooted binary tree $U_i$ has
$2n_i-1$ nodes, giving $2n_i$ choices for its coordinate
when the null symbol is included. Thus the number of possible
states is
\[
 N=\prod_{i=1}^{t}2n_i.
\]
By the arithmetic--geometric mean inequality, we have 
\[
 N
 \leq
 \left(\frac{\sum_{i=1}^{t}2n_i}{t}\right)^t
 =
 \left(\frac{2(n-1)}{t}\right)^t.
\]

Each state $\pi$ has at most $4^t$ transitions.
For a transition $(\pi_1,\pi_2)$, write $A=C(\pi_1)$,
$B=C(\pi_2)$, $O=S\setminus C(\pi)=S\setminus(A\cup B)$.
Here $A$ and $B$ are the two child leaf sets, and $O$ is the
set of outside taxa. Using membership tables for the state
leaf sets, the charge $c_Q(A,B,O)$ can be computed in~$O(n^4)$ 
time by scanning all four-taxon subsets and
processing each in constant time. The child values are
already available when states are processed in increasing
order of their leaf-set sizes. Therefore, each transition
can be evaluated in~$O(n^4)$ time, and the total running
time is
\[
 O\!\left(
 n^4\,4^t\prod_{i=1}^{t}2n_i
 \right).
\]
\end{proof}
\paragraph{Remark} 
In the end of the proof of Lemma~\ref{lem:merge}, 
substituting the bound on~$N = \prod_{i=1}^{t}2n_i$ gives
\[
 O\!\left(
 n^4\left(\frac{8(n-1)}{t}\right)^t
 \right),
\]
which retains the dependence on $n/t$ needed in the
subsequent analysis.

\subsection{The anchored algorithm}

Compute $p=\lfloor4k/n\rfloor$. For every anchor, enumerate the family in
Lemma~\ref{lem:color}, reconstruct all class trees, and run
\eqref{eq:product} for each surviving coloring. Accept and return a tree
if its score is at most $k$. Otherwise, reject if every branch fails.

\begin{proposition}\label{prop:anchored}
For $n\geq4$, the anchored algorithm is exact. With $h$ from Lemma~\ref{lem:color}, its time and space are at most
\[
 2^{O(h\log(1+n/h))}n^{O(1)}.
\]
\end{proposition}
\begin{proof}
Soundness follows from computing the full quartet score. On a yes-instance,
choose the anchor in Lemma~\ref{lem:anchor} and the coloring guaranteed by
Lemma~\ref{lem:color}. Each reconstructed $U_i$ equals the target rooted
restriction, so Lemma~\ref{lem:merge} returns a score at most $k$.
The family bound follows from Lemma~\ref{lem:color}. For each merge, AM--GM and $\sum_{i=1}^t n_i=m=n-1$ give
\[
 4^t\prod_i2n_i\leq(8m/t)^t.
\]
For fixed $m$, the function $x\log(1+8m/x)$ is increasing on~$x>0$. To see this, set $z=8m/x>0$ and 
observe that its derivative has the sign of $\log(1+z)-z/(1+z)>0$ for $z>0$. 
Since $t\leq 2h$, we have 
\[
\begin{aligned}
 (8m/t)^t&\leq(1+8m/t)^t \leq(1+4m/h)^{2h} =2^{O(h\log(1+m/h))}.
\end{aligned}
\]
This includes the injective case $t=m$ and $h\geq m$. Multiply by the family size, yet the exponents add and remain of the same order. 
The $n$ anchors contribute a polynomial factor. 
The $p=0$ case has $h=1$ and remains polynomial. 
Classes containing one or two leaves have a unique rooted
binary tree topology.
\end{proof}

\section{Entropy analysis and removal of the logarithm}
\label{sec:entropy}

\begin{lemma}\label{lem:entropy}
For all $h,n>0$,
\[
 h\log_2(1+n/h)\leq\frac{4}{\ln2}(nh^3)^{1/4}.
\]
\end{lemma}
\begin{proof}
For $u>0$, $u^{3/4}\leq1+u$. Therefore
$1/(1+u)\leq u^{-3/4}$, and integration gives
$\ln(1+x)\leq4x^{1/4}$. Substitute $x=n/h$ and multiply by $h/\ln2$.
\end{proof}

\begin{proof}[Proof of Theorem~\ref{thm:main}]
Handle $n\leq3$, $k\geq\binom n4$, and $n>4k$ in polynomial time as above.
Otherwise $p=\lfloor4k/n\rfloor\geq1$. The least-power-of-two choice of~$h$ gives
\[
 h<2p^{1/3},\,\text{ and }\; nh^3<8np\leq32k.
\]
Proposition~\ref{prop:anchored} and Lemma~\ref{lem:entropy} now bound the
anchored algorithm by $2^{O(k^{1/4})}n^{O(1)}$ time and space.
The exponents of all polynomial factors are constants independent
of~$h,k,n$. The additive statement follows, for example, from
$f(k)n^c\leq(f(k)^2+n^{2c})/2$, where $f(k)=2^{O(k^{1/4})}$.
Polynomial preprocessing and the trivial cases satisfy the same bound.
\end{proof}

\paragraph{Remark}
The bound $2^{O(h\log(1+n/h))}n^{O(1)}$ retains the
dependence on both $n$ and $h$. For a fixed error budget~$k$,
the guaranteed anchored error bound
$p=\lfloor 4k/n\rfloor$ is nonincreasing in~$n$, and hence
so is the chosen coloring parameter~$h$.
Moreover, when $n=\Theta(h)$, we have $h\log(1+n/h)=\Theta(n)$.
Thus both the coloring-family construction and the merge
computation run in $2^{O(n)}$ time in this regime, including
their polynomial overhead. Retaining the ratio $n/h$ inside
the logarithm gives this sharper estimate. A coarser
$n^{O(h)}$ bound would yield only $2^{O(n\log n)}$.

\section{Generic ETH consequences}
\label{sec:lower}

The Exponential Time Hypothesis (ETH) asserts a positive exponential-time
rate for 3-SAT in its number of variables~\cite{IP01}. Together with
sparsification~\cite{IPZ01}, it rules out a deterministic $2^{o(N)}$
algorithm on 3-CNF instances with $N$ variables and $O(N)$ clauses. The
following consequence uses the established \textsf{NP}-completeness of complete
MQI~\cite{GN03} to provide a generic observation. 

\begin{proposition}[Generic ETH barrier]\label{prop:genericeth}
Assuming ETH, complete MQI has no deterministic algorithm with running time
$2^{k^{o(1)}}n^{O(1)}$. Here the exponent means a function $g(k)$ such that,
for every fixed $\eta>0$, $g(k)=O((k+2)^\eta)$.
\end{proposition}
\begin{proof}
Assume ETH. By the sparsification lemma~\cite{IPZ01},
there is no deterministic $2^{o(N)}$-time algorithm for
3-SAT on instances with $N$ variables and $O(N)$ clauses.

Since complete MQI is \textsf{NP}-complete~\cite{GN03}, there exists
a polynomial-time many-one reduction from 3-SAT to complete
MQI. Let $\Phi$ be a 3-CNF formula with $N$ variables and
$O(N)$ clauses, and let $(Q,k)$ be the resulting complete
MQI instance on $n = |S|$ taxa. The reduction guarantees that
\begin{align*}
 &\Phi \text{ is satisfiable}
 \;\text{ if and only if } \\ 
 & (Q,k) \text{ is a yes-instance of complete MQI}.
\end{align*}
The encoding length of $\Phi$ is $O(N\log N)$.
Since the reduction runs in polynomial time, its output
has polynomial encoding length. Consequently, there is
a fixed constant $d\geq1$, depending only on the chosen
reduction, such that $n=O(N^d)$.

We may replace $k$ by $\min\{k,\binom{n}{4}\}$ without
changing the answer: every tree disagrees with at most
$\binom{n}{4}$ input quartets. Thus we may assume
\[
 0\leq k\leq\binom{n}{4}\leq n^4=O(N^{4d}).
\]

Suppose, for a contradiction, that complete MQI admits
a deterministic algorithm with running time
$2^{g(k)}n^{O(1)}$, where $g(k)=O((k+2)^\eta)$ for every fixed $\eta>0$.
Choose the fixed value $\eta=1/(8d)$. The preceding
parameter bound then gives
\[
 g(k)
 =O\!\left((k+2)^{1/(8d)}\right)
 =O(N^{1/2}).
\]
Therefore, applying the reduction and then this algorithm
would decide whether $\Phi$ is satisfiable in time
\[
 N^{O(1)}
 +2^{O(N^{1/2})}N^{O(1)}
 =2^{o(N)}.
\]
This contradicts the sparse-3-SAT consequence of ETH.
\end{proof}
In particular, this excludes parameter factors
$2^{O((\log(k+2))^c)}$ for every fixed $c>0$. It does not exclude
$2^{O(k^\delta)}$ for a specified fixed $\delta>0$, such as $\delta=1/5$,
and does not establish optimality of Theorem~\ref{thm:main}. 
The argument uses only the existence of a polynomial-time
reduction and does not determine its degree explicitly.
It therefore establishes the generic barrier above,
without identifying a specific constant $\delta>0$ for
which a running time of $2^{o(k^\delta)}n^{O(1)}$ is
excluded under ETH.

\paragraph{Remark} 
A more precise analysis of the reduction could yield a
numerical ETH lower bound. Specifically, suppose a
deterministic polynomial-time reduction from sparse 3-SAT
with $N$ variables produces complete MQI instances with
$n=N^{O(1)}$ taxa and budget $k=O(N^\alpha)$ for a fixed
$\alpha>0$. A hypothetical
$2^{o(k^{1/\alpha})}n^{O(1)}$ algorithm would then solve
sparse 3-SAT in $2^{o(N)}$ time, contradicting ETH.
Determining an explicit value of $\alpha$ therefore
requires a quantitative analysis of the reduction.

The reduction chain analyzed by Dupr\'e la Tour, Lafond, and
Ndiaye~\cite[proof of Theorem~5]{DLN26} concerns general
Unrooted Quartet Consistency, whose input need not be complete.
Transferring this approach to complete MQI requires a reduction
that preserves yes- and no-instances and explicitly bounds
both the number of taxa and the disagreement budget.
Arbitrarily assigning topologies to unspecified four-sets can
destroy compatibility, and complete MQI with $k=0$ is
polynomial-time solvable~\cite{ESSW99}.
Thus, their reduction does not directly establish a
parameterized lower bound for complete MQI.
We leave open whether complete MQI admits an algorithm with
running time $2^{o(k^{1/4})}n^{O(1)}$.

\section{Concluding remarks}
\label{sec:concluding}

Low-error anchoring and size-sensitive counts yield a deterministic
$2^{O(k^{1/4})}+n^{O(1)}$ algorithm for complete unweighted MQI.
The algorithm recovers rooted restrictions from anchored triplets and
merges them using the full quartet objective. Retaining the number of
repair subsets and the actual class sizes is essential to the logarithm-free
analysis.

The generic ETH consequence in Section~\ref{sec:lower} does not establish
optimality of the fourth-root exponent. A matching lower bound requires
a reduction that preserves completeness and controls the error budget.
Establishing such a reduction, obtaining a polynomial-space variant,
and improving the explicit constants and polynomial factors are natural
directions for further work.

\section*{Declaration of AI assistance}
During the preparation of this work the author used ChatGPT 6.1 in order to 
perform preliminary literature survey and polish the writing and readability. 
After using this tool/service, the author reviewed and edited 
the content as needed and takes full responsibility for the content of the published article.

\bibliographystyle{elsarticle-num-names}
\bibliography{anchor_references}
\end{document}